\documentclass[prl,twocolumn,showpacs,floatfix]{revtex4}
\usepackage{graphicx}

\def\dderiv#1#2{\frac{\partial^2#1}{\partial#2^2}}
\def\xderiv#1#2#3{\frac{\partial^2#1}{\partial#2\partial#3}}

\def\nderiv#1#2#3{\frac{\partial^#1#2}{\partial#3^#1}}
\def\eqref#1{(\ref{#1})}

\begin{document}

\title {The stability of solitons in biomembranes and nerves}
\author{B. Lautrup, A. D. Jackson and T. Heimburg}
\affiliation{Niels Bohr Institute, Blegdamsvej 17, DK-2100,
Copenhagen \O, Denmark}
\date{\today}

\begin{abstract}

We examine the stability of a class of solitons, obtained from a
generalization of the Boussinesq equation, which have been proposed
to be relevant for pulse propagation in biomembranes and nerves.
These solitons are found to be stable with respect to small
amplitude fluctuations.  They emerge naturally from non-solitonic
initial excitations and are robust in the presence of dissipation.

\end{abstract}

\pacs{80,87,87.10}

\maketitle

\section{Introduction}

The action potential in nerves is a propagating voltage
pulse across the axonal membrane with an amplitude of about 100\,mV. In
1952, A. L. Hodgkin and A. F. Huxley proposed a theory for the nerve
pulse which has since become the textbook model \cite{Hodgkin52}.
Their picture is based on the equilibration of ion gradients across the
nerve membrane through specific ion-conducting proteins (called ion channels)
which leads to transient voltage changes.  Hodgkin-Huxley theory thus
relies on dissipative processes and is intrinsically not isentropic.
It is rather based on Kirchhoff circuits involing capacitors (the
nerve membrane), resistors (the ion channels) and electrical currents
introduced by the ion fluxes.

We have recently proposed an alternative model for the nerve pulses
based on the propagation of a localized density pulse (soliton) in
the axon membrane \cite{HJ}.  This model has several advantages over
the Hodgkin-Huxley model.  It explains the reversible temperature
and heat changes observed in connection with the nerve pulse. (Such
reversible changes are not consistent with the Hodgkin-Huxley theory
but rather suggest that an isentropic process is responsible for the
action potential \cite{Abbott58,Howarth68,Ritchie85}.)  It further
predicts the correct pulse propagation velocities in myelinated
nerves.  These velocities are closely related to the lateral sound
velocities in the nerve membrane.  One essential feature of our
model is the presence of empirically known lipid phase transitions
slightly below physiological temperatures.  The closer the phase
transition is to physiological temperatures, the easier it is to
excite the nerve pulse.  Our model therefore immediately explains
another interesting feature of nerve excitation, i.e., that the
nerve pulse can be induced by a sudden cooling of the nerve, and
that it can be inhibited by a temperature increase
\cite{Kobatake71}.  During compression, the appearance of a voltage
pulse is merely a consequence of the piezo-electric nature of the
nerve membrane, which is partially charged and asymmetric.

Another advantage of a soliton-based description of pulse
propagation in nerves lies in its predictive power.  Given measured
values of the compression modulus as a function of lateral density
and frequency, soliton properties (including its shape and its
energy) can be determined uniquely as a function of soliton
velocity. Given a measured soliton velocity, the theory contains no
freely adjustable parameters and has the virtue of being
falsifiable.

In \cite{HJ} the possibility of soliton propagation was explored and
compared to observations in real nerves.  In the present paper we study
some intrinsic features of these solitons, in particular the stability
of such pulses in the presence of noise and dissipation.  Such
investigations are necessary to demonstrate that such pulses could
propagate under realistic physiological conditions over the length
scales of nerves (as much as several meters) even in the presence
of friction and lateral heterogeneities.  In the following
section, we will state the model more precisely and derive the analytic
form of its solitonic solutions.  We will then turn to a description of
the numerical methods used here.  We use these methods to probe (i) the
stability of solitons with respect to ``infinitesimal'' perturbations
(i.e., lattice noise), (ii) the way in which solitons are produced by localized
non-solitonic initial excitations of the system, and (iii) the
behavior of solitons in the presence of dissipation.  We will demonstrate
that the solitons of Ref.\,\cite{HJ} are remarkably robust with respect to all
of these perturbations.

\section{Analytic Considerations}

Thermodynamic measurements of the lipids of biological membranes
reveal a number of interesting features of potential relevance for
understanding the nature of pulses in biomembranes and nerves.  In
particular, such systems display an order-disorder transition at
temperatures somewhat below that of biological interest from a low
temperature ``solid-ordered'' phase to a high temperature
``liquid-disordered'' phase in which both the lateral order and
chain order of the lipid molecules is lost \cite{Ipsen87}.  The
proximity of this phase transition to temperatures of biological
interest has striking effects on the compression modulus and, hence,
on the sound velocity \cite{Heimburg98,Halstenberg98}.  For
densities some $10$\% above the equilibrium density, the
low-frequency sound velocity is reduced by roughly a factor of $3$
from the velocity of $c_0 = 176.6$\,m/s found at equilibrium.  The
sound velocity then rises sharply, returning to the value $c_0$ at a
density roughly $20$\% above the equilibrium density.  Measurements
at high frequencies (i.e., $5$\,MHz) reveal  a much smaller dip in
the lateral compression modulus and a sound velocity that is always
materially larger than that at low frequencies and thus indicate the
presence of significant dispersion \cite{Halstenberg98,Schrader02}.

In Ref.\,\cite{HJ}, these features were exploited to suggest that the
propagation of sound in these lipid mixtures can be described by the
equation
\begin{eqnarray}\label{ur1}
\frac{\partial^2}{\partial \tau^2} \Delta \rho^A & = & \frac{\partial}{\partial z} \left[
\left( c_0^2 + p \Delta \rho^A + q (\Delta \rho^A )^2 \right) \frac{\partial}{\partial z}
\Delta \rho^A \right] \nonumber \\
{} & {} &   - h \frac{\partial^4}{\partial z^4} \Delta \rho^A \ .
\end{eqnarray}
Here, $\Delta \rho^A = \rho^A - \rho_0^A$ is the difference between
the lateral mass density of the membrane and its empirical
equilibrium value of $\rho^A_0 = 4.035 \times 10^{-3}$\,g/m$^2$, and
the low frequency sound velocity is $c_0 = 176.6$\,m/s.  The
coefficients $p$ and $q$ were fitted  to measured values of the
sound velocity as a function of density.  Although high frequency
sound velocity measurements indicate that the dispersive
coefficient, $h$, must be positive, neither the magnitude of $h$ nor
the specific form of this term have been verified experimentally. In
practice, the only role of $h$ is to establish the linear size of
solitons, and it can thus be chosen, e.g., so that the width of the
soliton is comparable to that known for nerve pulses.  Here, we
choose to work with the dimensionless variables $u$, $x$ and $t$
defined as
\begin{equation}\label{ur2}
u = \frac{\Delta \rho^A}{\rho^A_0} \ \ x = \frac{c_0}{\sqrt{h}} z \
\ t = \frac{c_0^2}{\sqrt{h}} \tau  \ .
\end{equation}
With this choice of variables, eq.\,(\ref{ur1}) assumes the form
\begin{equation}\label{eEqMotion}
\frac{\partial^2 u}{\partial t^2} = \frac{\partial}{\partial x}
\left( B(u) \frac{\partial u}{\partial x} \right) - \frac{\partial^4
u}{\partial x^4}
\label{1}
\end{equation}
with
\begin{equation}
B(u) = 1 + B_1 u + B_2 u^2 \ .
\label{2}
\end{equation}
The qualitative features of the empirical compression modulus
require that $B_1 < 0$ and $B_2 > 0$.  In the numerical work described
below, we will adopt the parameter values $B_1 = -16.6$ and $B_2 = 79.5$
found in Ref.\,\cite{HJ}.  Eq.\,(\ref{1}) can be recognized as a
generalization of the Boussinesq equation, and it is known to
have exponentially localized ``solitonic'' solutions which propagate
without distortion for a finite range of sub-sonic velocities.  We
now determine the analytic form of these solitons.

Since we seek solutions which propagate without distortion, we
regard $u$ as a function of $\xi = x - \beta t$ and rewrite
Eq.\,(\ref{1}) as
\begin{equation}
\beta^2 \frac{\partial^2 u}{\partial \xi^2} =
\frac{\partial}{\partial \xi} \left( B(u ) \frac{\partial
u}{\partial \xi} \right) - \frac{\partial^4 u}{\partial \xi^4} \ .
\label{3}
\end{equation}
We can integrate this equation twice  with the assumption that $u$
vanishes at spatially infinity to obtain
\begin{equation}
\frac{\partial^2 u}{\partial \xi^2} = (1 - \beta^2 )u + \frac{1}{2}
B_1 u^2 + \frac{1}{3} B_2 u^3 \ . \label{4}
\end{equation}
It is clear from this equation that exponentially localized
solutions are possible if $\beta^2 < 1$.  Multiplication by
$\partial u / \partial \xi$ and a final integration leave us with
the result
\begin{equation}
\left( \frac{\partial u}{\partial \xi} \right)^2 = (1 -\beta^2 ) u^2
+ \frac{1}{3} B_1 u^3 + \frac{1}{6} B_2 u^4 \ . \label{5}
\end{equation}
It is clear that $u$ is symmetric about its maximum value. The
solution will grow from $0$ until it reaches a maximum value at
which $\partial u / \partial \xi = 0$. Equation (\ref{5}) indicates
that this is possible only if
\begin{equation}
1 > | \beta | > \beta_0 = \sqrt{1 - \frac{B_1^2}{6 B_2}} \ . \label{6}
\end{equation}
For the parameters $B_1=-16.6$ and $B_2=79.5$ adopted in \cite{HJ},
we find $\beta_0 \approx 0.649851$.  We will use these parameter
values in the remainder of this paper.

\begin{figure}[ht]
\includegraphics[width=\hsize]{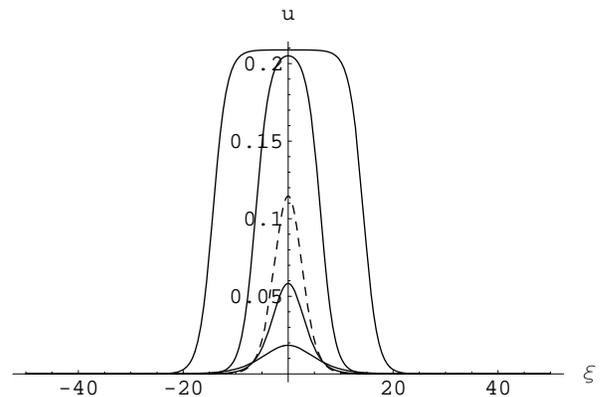}
\caption{Soliton profiles for velocities $\beta =
\beta_0+4\times10^{-9}$, $0.65$, $0.734761$, $0.85$, and $0.95$. The
maximum height diminishes as a function of $\beta$.  The width of
the soliton diverges for both $\beta \to \beta_0$ and $\beta \to 1$
and has a minimum at $\beta \approx 0.734761$, which corresponds to
the dashed curve.} \label{fSolitons}
\end{figure}

We thus expect localized solutions for $\beta_0 < | \beta | < 1$.  When
this condition is met, the right side of Eq.\,(\ref{5}) will have
two real roots, $u = a_{\pm}$ with
\begin{equation}
a_{\pm} = -\frac{B_1}{B_2} \left( 1 \pm \sqrt{\frac{\beta^2 -
\beta_0^2}{1 - \beta_0^2}} \right) \ . \label{7}
\end{equation}
It is readily verified that the  desired solitonic solutions of
Eq.\,(\ref{1}) have the analytic form
\begin{equation}
u (\xi )= \frac{2 a_+ a_-}{(a_+ + a_-) + (a_+ - a_-)\cosh{\left( \xi
\sqrt{1 - \beta^2} \right)}} \ . \label{8}
\end{equation}
These solutions are shown in Figure \ref{fSolitons} for a selection
of soliton velocities.

As expected, Eq.\,(\ref{1}) can be obtained from a suitable energy
density. We thus seek an energy density, ${\cal E}$, such that
Eq.\,(\ref{1}) will result from variation of the corresponding
Lagrangian density. To this end, it is useful to introduce the
dimensionless displacement, $s(x,t)$, defined as $u =
\partial s /
\partial x$. The energy density can then be written as
\begin{equation}
{\cal E} = \frac{1}{2} \left( \frac{\partial s}{\partial t}
\right)^2 + \left[ \frac{1}{2} u^2 A(u) + \frac{1}{2} \left(
\frac{\partial u}{\partial x} \right)^2 \right] \label{9}
\end{equation}
with
\begin{equation}
A(u) = 1 + \frac{1}{3} B_1 u + \frac{1}{6} B_2 u^2 \ . \label{10}
\end{equation}
The two terms in Eq.\,(\ref{9}) represent the kinetic and potential
energy densities, respectively.  The corresponding Lagrangian
density is obtained by changing the sign of the potential energy term,
and Eq.\,(1) follows by standard variational arguments.  This form of the
energy density leads to two important observations.  First, we note
that the energy density simplifies considerably if $u$ describes a
soliton and is given by Eq.\,(\ref{8}).  Specifically, use of the equation
of motion allows us to write the energy density as ${\cal E}_{\rm sol}
= u^2 A(u)$.  The specific form of Eq.\,(\ref{8}) is sufficiently simple that the
energy of a soliton can be calculated analytically and involves only
elementary functions.

It is also useful to consider the  total energy associated with an
arbitrary solution, $u(x,t)$, of Eq.\,(\ref{1}) as given by the
integral over all space of the energy density, ${\cal E}$, of
Eq.\,(\ref{9}).  Recognizing perfect differentials when they arise
and making use of the equation of motion, Eq.\,(\ref{1}), we find
the expected result that the energy is independent of time for an
arbitrary choice of $u(x,t)$.  (This result assumes either that
$u(x,t)$ vanishes as $|x| \to \infty$ or that it satisfies periodic
boundary conditions in $x$.)  It is also useful to consider the time
dependence of the the integral of $u$ over all space,
\begin{equation}
U=\int u(x,t)\,dx~.
\end{equation}
It is clear from the equation of motion that $\partial^2 U /\partial
t^2$ can be expressed as an integral of perfect differentials.
Hence, $\partial^2 U / \partial t^2 = 0$ if $u$ vanishes at spatial
infinity or is periodic.  Thus, the time dependence of $U$ is
elementary and can include only a constant term and a term linear in
$t$.  As we shall see below, $U$ is independent of time when
$u(x,t)$ is periodic.

\section{Numerical considerations}

We would like to investigate a number of questions associated with
the stability of the solitons of Eq.\,(\ref{8}).  Although the
simplicity of the analytic form of these solitons suggests that it
may be possible to solve the problem of infinitesimal stability
analytically, we have elected to consider this problem numerically.
To this end, it is convenient to re-write Eq.\,(\ref{1}) as two first-order
equations.  We obtain
\begin{equation}
\frac{\partial u}{\partial t} = \frac{\partial v}{\partial x}~,~~~~
\frac{\partial v}{\partial t} = \frac{\partial f}{\partial x}
\label{11}
\end{equation}
with
\begin{equation}
f= u + \frac{1}{2} B_1 u + \frac{1}{3} B_2 u^2 - \frac{\partial
w}{\partial x} \ , \label{12}
\end{equation}
where $w = \partial u / \partial x$ (and incidentally $v=\partial
s/\partial t$). (Note that the first of Eqs.\,(\ref{11}) ensures
that the spatial integral of $u$ is independent of time if $v$ is
chosen to be periodic.) Equations (\ref{11}) are well-suited to
numerical solution using a variant of the two-step Lax-Wendroff
method \cite{NR}.  We consider the function, $u(x,t)$, on a primary
mesh of equally spaced points, $(p \Delta x , q \Delta t)$, where it
has the values $u_{pq}$. (Evidently, we must demand $\Delta x <
\beta \Delta t$ in order to satisfy the usual Courant condition,
which is necessary but not sufficient to ensure numerical
stability.) The mesh is then extended to include half-integer values
of $p$ and/or $q$. It is evident from Eqs.\,(\ref{11}) and
(\ref{12}) that the functions $u$, $v$, and $f$ live naturally on
the points $(p,q)$ and $(p + 1/2,q+1/2)$ and that $w$ lives
naturally on the points $(p+1/2,q)$ and $(p,q+1/2)$.

The algorithm is implemented as follows:  Establish the initial
values of $u$ and $v$ on the points $(p,0)$.  (If $u$ is a solitonic
solution, we simply have $v(x,0)=-\beta u(x,0)$.)  At every time
step, the values of $w_{p+1/2,q}$ are obtained as
\begin{equation}
w_{p+1/2,q} = \frac{u_{p+1,q}-u_{p,q}}{\Delta x} \ , \label{13}
\end{equation}
and these values can be used to construct $f_{pq}$  using a
similarly symmetric difference formula.  One then proceeds to the
evaluation of the functions $u$ and $v$ a time $\Delta t/2$ later.
In proceeding from $q$ to $q + 1/2$, we define,
\begin{equation}
u_{p+1/2,q+1/2} = \frac{1}{2} \left( u_{p,q} + u_{p+1,q} \right) +
\frac{\Delta t}{2 \Delta x} \left( v_{p+1,q} - v_{p,q} \right) \ .
\label{14}
\end{equation}
The values of $w_{p,q}$ and $f_{p+1/2,q+1/2}$  are obtained as in
Eq.\,(\ref{13}), and $v_{p+1/2,q+1/2}$ is obtained as in
Eq.\,(\ref{14}).  A slightly different procedure is adopted in going
from time $q +1/2$ to $q+1$. In this case,
\begin{equation}
u_{p,q+1} = u_{p,q}+ \frac{\Delta x}{\Delta t} \left(
v_{p+1/2,q+1/2} - v_{p-1/2,q+1/2} \right) \, , \label{15}
\end{equation}
the values of $w_{p+1/2,q+1}$ and $f_{p,q+1}$ are obtained as in
Eq.\,(\ref{13}), and $v_{p,q+1}$ is obtained in analogy with
Eq.\,({15}).  This algorithm is both fast and stable in practice.
(For the periodic boundary conditions and the choice of $\Delta x =
0.1$ and $\Delta t = 0.001$, used below, it was possible to follow
$10^6$ time steps without discernible loss of accuracy.)

It is useful to note that energy of Eq.\,(\ref{9}) is not rigorously
conserved by this numerical algorithm.  In the following numerical
examples, the energy was found to decrease at a roughly constant
rate proportional to $\Delta x^2$.  This fact was used to make an
appropriate choice of $\Delta x = 0.1$.  The corresponding value of
$\Delta t = 0.001$ was selected to yield full numerical stability.

\section{Numerical results}

\subsection{Small amplitude noise}

Our primary numerical concern is to study the stability of the
solitonic solutions of Eq.\,(\ref{8}) with respect to
``infinitesimal'' perturbations.  We employ the parameters $B_1 =
-16.6$ and $B_2 = 79.5$ adopted in \cite{HJ}, for which $\beta_0
\approx 0.650$. We will show results for an initial soliton with
velocity $\beta = \beta_1 \approx 0.735$. This soliton has a width
(i.e., full width at half maximum) of roughly $6.24$, which is the
minimum width possible for the values of $B_1$ and $B_2$ considered.
There is, of course, no reason to believe that a soliton on a
discrete lattice with finite $\Delta x$ will have a profile
identical to the analytic form of Eq.\,(\ref{8}).  The use of this
analytic form in establishing the initial values of $u$ and $v$ thus
inevitably introduces a measure of noise into the numerical system.
Since there is no other ``natural'' choice for the initial form of
the solitonic excitation, this noise represents the best
approximation to infinitesimal perturbations that can be realized in
a numerical study.

In an analytic approach to the question of infinitesimal stability,
one considers the time evolution of the sum of the soliton under
investigation and a small excitation, $\delta u(x,t)=\psi(x,t)$. The
equation of motion \eqref{eEqMotion} is then expanded to first order
in $\psi$, and expressed in terms of $t$ and $\xi=x-\beta t$,
\begin{equation}
\dderiv\psi t-2\beta\xderiv\psi t\xi+\beta^2\dderiv\psi\xi
=\dderiv{(B(u)\psi)}\xi-\nderiv4\psi\xi \, .
\end{equation}
It follows that solutions to this (non-Hermitean) equation can be
written as the product of functions $\psi_\lambda(\xi)$ and
$\exp{(\lambda t)}$.  If one or more of the resulting values of
$\lambda$ has a positive real part, the corresponding
$\psi_{\lambda} (x,t)$ will grow exponentially with time, and the
initial solitonic solution will be locally unstable.  Since it is
our aim to detect precisely such exponential stabilities (if
present), it is of no consequence that the numerical noise
introduced by the finite mesh size is small.  Exponential
instabilities will be apparent if they are present.  The finite size
of $\Delta x$ also means that there is a smallest wave length
perturbation which can be studied on the lattice. In practice,
potential instabilities involving such wave lengths will be
invisible to numerical studies only if they are orthogonal to those
wave lengths which \textit{can} be investigated reliably with the
$\Delta x$ chosen.  While this is not impossible, it is unlikely.

\begin{figure}[ht]
\includegraphics[width=\hsize]{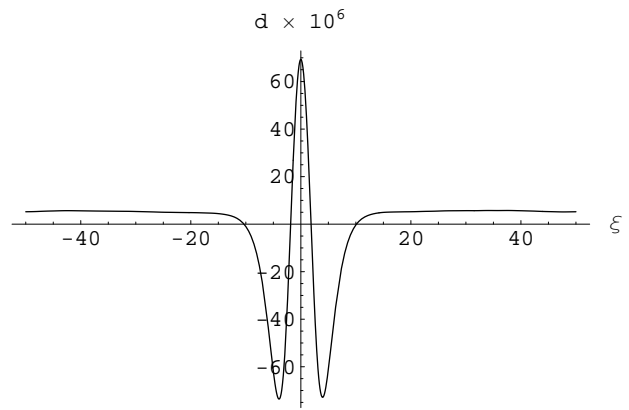}
\caption{The difference $d(\xi)$ between the time-averaged numerical
soliton and the analytic soliton for the minimum width soliton with
$\beta=\beta_1\approx0.735$. The average has been performed over
1000 units of time, during which the soliton travels more than 100
times its own width.}
 \label{fAveDif}
\end{figure}

Results were obtained with $\Delta x = 0.1$ and $\Delta t = 0.001$.
The spatial lattice was chosen to be periodic with length $100$. For
$\beta =\beta_1\approx 0.735$, the exact energy of the soliton is
$0.0377$. The energy of this initial state is smaller by $1.5 \times
10^{-6}$ when calculated on the lattice.  Energy is not strictly
conserved by the numerical algorithm adopted but rather decreases
linearly with time over the time intervals considered.  In the
present case, energy is lost at the rate of $7.3 \times 10^{-9}$ per
unit  time.  We have followed this soliton for times as long as
$1000$ units, during which the soliton can propagate more than $100$
times its own width. The energy loss is negligible, and there is
absolutely no indication of instability.  (Note that the discrepancy
in the initial energy is proportional to $\Delta x^2$; the rate of
energy loss scales like $\Delta x^3$.)

We can illustrate soliton stability in the following manner.  We
first determine the location of the maximum of the soliton as a
function of time.  The constancy of its velocity over large time
intervals provides an initial indication of the stability of the
soliton.  In the present case, this velocity is found to be stable
roughly $0.02$\% less than the initial velocity of the analytic
soliton.  (This error scales with $\Delta x^2$.)  There are, of
course, small fluctuations in the location of both the maximum
density and, hence, the velocity due to the presence of noise. For
the present example, such fluctuations in the location of the
maximum are never greater than $0.004$, which is $25$ times smaller
than $\Delta x$. (These fluctuations also scale like $\Delta x^2$.)
Having identified the position of soliton as a function of time,
each time frame is shifted to locate the soliton at a common point.
A time-averaged soliton is then constructed in order to minimize the
effects of noise.   The difference between the time-averaged soliton
and the analytic soliton is shown in Figure \ref{fAveDif}.  The peak
value of the time-averaged soliton is slightly (i.e., roughly
$0.05$\%) higher than that of the analytic soliton, and it is
somewhat narrower than its analytic counterpart. (The size of these
differences again scales with $\Delta x^2$.) This demonstrates the
claim that the analytic solitons are not identical to solitons on a
finite mesh. Further, the systematic discrepancy between these two
solutions is the source of and has a magnitude comparable to that of
the noise in the system.

\begin{figure}[ht]
\includegraphics[width=\hsize]{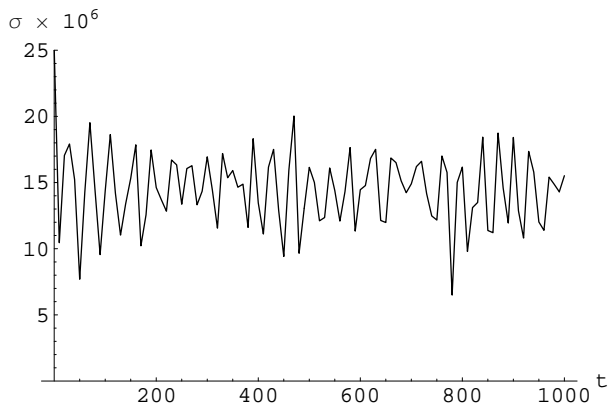}
\caption{Time evolution of rms noise level $\sigma$ for the minimal
width soliton over 1000 units of time. }
 \label{fNoise}
\end{figure}

We now consider the nature of the ``lattice noise'' in the system as
a function of time by subtracting the time averaged soliton from the
full $u(x,t)$ at each time update and constructing the root mean
square of the resulting noise as a function of time.  If the soliton
is stable, the  resulting rms noise should be bounded as a function
of time.  If the soliton were unstable, however, we would expect to
find systematic differences in the vicinity of the soliton maximum
which are well above noise level and which grow exponentially with
time. The spatial distribution of noise at later times shows no sign
of such systematic effects, and its magnitude is the same both near
and far from the location of the soliton.  The calculated rms noise
is shown in Figure \ref{fNoise} as a function of time.  Again, there
is no sign of such instabilities.  Since qualitatively similar
results are found for other values of $\beta$, we conclude that the
solitons of Eq.\,(\ref{1}) are stable with respect to small
perturbations.

It is also possible to study soliton stability in the presence of
larger amplitude noise.  This is most easily done by choosing a form
of $u(x,0)$ which consists of both the (analytic) soliton of
interest and a linear combination of the lowest $k\le K$ periodic
waves on the interval $L$, $a_k \sin{(2\pi k x/ L + \phi_k)}$, with
phases chosen at random and amplitudes chosen at random subject to a
constraint on the overall rms noise level at $t=0$.  The analysis
proceeds as above. We have considered the case of $K=10$ with an
initial rms noise as large as $5$\% of the maximum amplitude of the
soliton. The results are similar to those found for small amplitude
noise: There are no indications of soliton instability.

\subsection{Soliton genesis}

\begin{figure}[ht]
\includegraphics[width=\hsize]{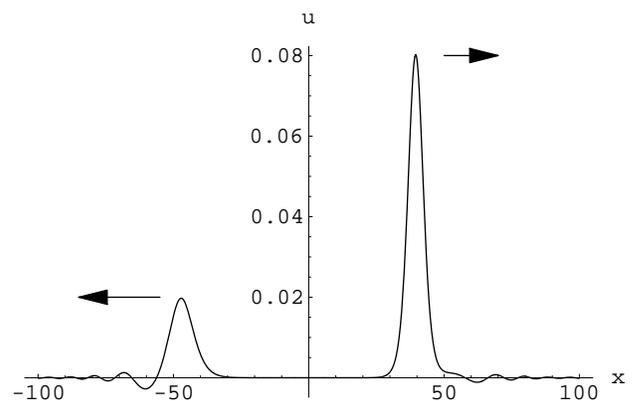}
\caption{A minimal width soliton with an initial velocity, $\beta$,
50\% lower than the corresponding analytic value, shown at $t=50$.
It has divided into two solitons of different sizes, propagating in
opposite directions.  Small-amplitude waves run ahead of the
solitons with velocity $\beta \approx 1$; the region between the
solitons is essentially noise free.  (See also Figure
\ref{fDisturb1}).  Note that the length of the periodic lattice has
been increased here to avoid interference effects between the
solitons and the leading small amplitude waves.} \label{fDisturb}
\end{figure}

\begin{figure}[ht]
\includegraphics[width=\hsize]{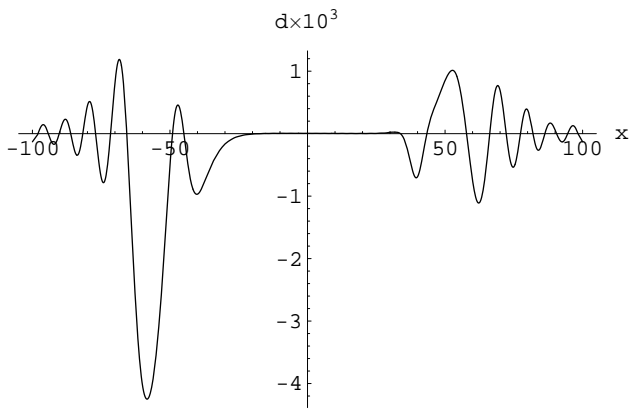}
\caption{The graph in Figure \ref{fDisturb} with the two solitons
subtracted out to leave only the small-amplitude waves running ahead
of the solitons.}
 \label{fDisturb1}
\end{figure}

It is also instructive to consider finite-amplitude disturbances and
to see how a localized but non-solitonic initial state evolves with
time. To illustrate this, we choose $u(x,0)$ to be the minimum width
soliton of Eq.\,(\ref{8}).  In this case, however, we distort the
second initial condition and choose $v(x,0) = - p\,\beta  u(x,0)$
with $p=0.5$.  Thus, the initial field is \textit{not} solitonic.
The time evolution shows that this initial pulse ``sheds'' matter
and changes its shape through the emission of a smaller soliton,
which moves in the opposite direction, and small amplitude waves,
which run ahead of the solitons with velocity $\beta \approx 1$. The
two solitons are captured in Figure \ref{fDisturb} at $t=50$. The
velocity of the larger soliton is $\beta=0.799$ and its maximum is
at $x=139.515$ whereas the smaller has $\beta=-0.948$ and maximum at
$x=52.871$.  The shape of each of these solitons is accurately
described using Eq.\,(\ref{8}) with the corresponding measured
velocity. These two solitons account for virtually all of the
initial energy of the system; approximately $0.3$\% of this energy
is associated with the small amplitude motion distinct from the
solitons. In Figure \ref{fDisturb1} the two solitons have been
subtracted out, and only the difference is plotted.  This confirms
that the shapes are indeed solitonic.

Similar results have been obtained for other non-solitonic initial
pulse forms (e.g., Gaussian pulses).  In short, for the cases
explored, non-solitonic initial excitations evolve into solitons and
small amplitude non-solitonic disturbances. In infinite space,
dispersion ensures that the solitonic and non-solitonic components
will become spatially distinct and that the amplitude of the latter
will decrease with time.  This is obviously not the case for the
periodic lattice considered here.

\subsection{Solitons and dissipation}

It is also possible to consider the consequences of dissipation on
soliton propagation.  The inclusion of viscosity in the
Navier-Stokes velocity results in an additional term on the right of
Eq.\,(\ref{1}) of the form $\kappa \partial^3 u / \partial x^2
\partial t$.  This term is readily incorporated in our numerical approach by the inclusion of the term
$+ \kappa \partial v /\partial \xi$ in Eq.\,(\ref{12}).  We have
performed numerical studies with the value $\kappa = 0.05$.  With
this choice of $\kappa$, the height of the soliton is reduced by
roughly $70$\% at $t=990$, has travelled more than 100 times its
initial width.  As energy is dissipated, the soliton accelerates,
and its profile changes with the expected drop in its amplitude.
Over the entire time range considered, we find that the soliton
profile is consistent with the analytic soliton profile of
Eq.\,(\ref{8}) appropriate for the corresponding instantaneous
velocity of the pulse.  This is illustrated in Figure 6, which shows
the comparison of analytic solitons (in infinite space) and these
numerical results including dissipation at several times.

\begin{figure}[ht]
\includegraphics[width=\hsize]{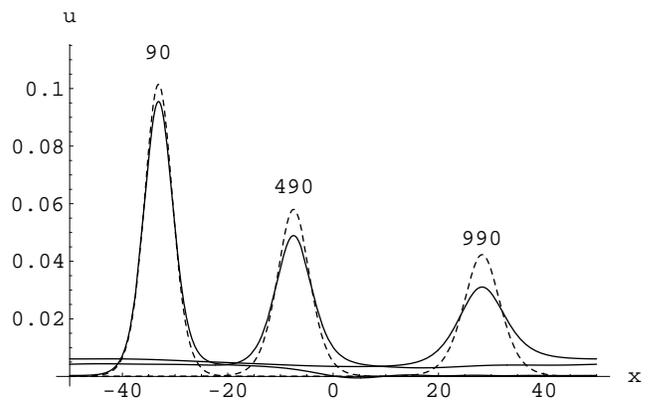}
\caption{Decaying soliton (fully drawn) with $\kappa=0.05$,
initially at $x=0$. The dashed curves depict the analytic solitons
with the instantaneous velocity of the numeric solitons. The numbers
above the peaks indicate the running time, and their particular
values have been chosen for illustrative purposes. The soliton has
in fact wrapped around the periodic lattice more than 9 times during
the time interval of the simulation.}
 \label{fDisturb2}
\end{figure}

For several reasons, this agreement is necessarily only approximate.
First, some time is required for the soliton profile to adjust to
the exact form corresponding to its instantaneous velocity.
Obviously, only a limited time is available for this adjustment in
the presence of dissipation.  More importantly, the
time-independence of the spatial integral of $u$ is not affected by
the inclusion of dissipation.  Thus, $u(x,t)$ approaches a constant
value for all $x$ as $t \to \infty$.  On a periodic lattice, as
here, this constant is non-zero.  This effect is clearly seen in
Figure 6, and it is obviously not included in the analytic form of
Eq.\,(\ref{8}) valid in infinite space. Figure 6 shows no indication
of the catastrophic break-up of the soliton into small amplitude
waves which might be anticipated in the presence of strong
dissipation. It should be noted that magnitude of the dissipation
considered here is large compared to what is to be expected in
biomembranes and nerves, where little or no change in pulse shape is
observed over distances roughly $20$ times than the pulse width.

\section{Conclusions}

We have considered here a number of tests of the stability of the
solitons associated with the modified Boussinesq equation,
Eq.\,(\ref{1}).  After finding the analytic form of these solitons,
we turned to a numerical investigation (with periodic boundary
conditions) of the their stability with respect to various
perturbations.  These solitons were found to be stable with respect
to the ``smallest possible'' perturbations inevitably induced by the
finite size of the numerical mesh and to finite but small periodic
perturbations.  Solitons are found to be produced by arbitrary
localized but non-solitonic initial excitations.  Finally, we have
shown that solitons retain their characteristic properties even in
the presence of relatively strong dissipation.   It was argued in
Ref.\,\cite{HJ} that the measured compression modulus of lipids of
biological membranes are suitable for the production of solitons.

These findings may be of immediate relevance for the
propagation of the action potential in nerve axons \cite{HJ}.  The
solitons described above are subject to friction and dissipation.
Nerve membranes are not homogeneous, i.e., they vary both in thickness
(e.g., at the site of the soma) and in the specific composition of
lipids and proteins.  Elastic constants may therefore vary locally.
In the present paper we have shown that neither noise nor dissipation
affect the propagation of solitary waves as such but rather lead only
to slight changes in amplitude and velocity.  These pulses are therefore
likely to be robust with respect to the unavoidable variance in shape and
composition of biological membranes and to dissipative hydrodynamic
processes which accompany the observed thickness changes in nerves
\cite{Iwasa80}.  Thus, the present results suggest that a model of
nerve pulses as {\em stable\/} solitons is viable even in a realistic
physiological environment and and that such a model may provide
an immediate and reliable explanation of associated mechanical \cite{Iwasa80}
and thermodynamic \cite{Abbott58,Howarth68,Ritchie85} effects
that remain unexplained in the presently accepted Hodgkin-Huxley model \cite{Hodgkin52}.

\acknowledgements{We thank Hans Fogedby, Mogens H{\o}gh Jensen,
Bo-Sture Skagerstam, and Erwin Neher for valuable discussions.}

\end{document}